\documentclass[useAMS,usenatbib]{mn2e}

\usepackage{graphicx}

\title[A class of relativistic stars with a linear equation
of state]{A class of relativistic stars with a linear equation of state}
\author[R. Sharma and S. D. Maharaj]{R. Sharma\thanks{E-mail:
206526115@ukzn.ac.za} and S. D. Maharaj\thanks{maharaj@ukzn.ac.za}\\
Astrophysics and Cosmology Research Unit\\
School of Mathematical Sciences\\
University of KwaZulu-Natal, Private Bag X54001, Durban 4000, South Africa.}

\begin{document}

\date{}

\pagerange{\pageref{firstpage}--\pageref{lastpage}} \pubyear{}

\maketitle

\label{firstpage}

\begin{abstract}
By assuming a particular mass function
we find new exact solutions to the Einstein field
equations with an anisotropic matter distribution.
The solutions are shown to be relevant for the description
of compact stars.  A distinguishing feature of this class of
solutions is that they admit a linear equation of state which
can be applied to strange stars with quark matter.
\end{abstract}


\section{Introduction}

In studies of compact relativistic astrophysical objects, strange
stars have long been proposed as an alternative to neutron star
models. The theoretical basis of strange stars composed of
deconfined $u$, $d$ and $s$ quarks is a direct consequence of the
conjecture that quark matter might be the true ground state of
hadrons \citep{Witten, Farhi}. Strange stars are expected to form
during the collapse of the core of a massive star after the
supernova explosion. Another possibility is that a rapidly spinning
neutron star can accrete sufficient mass to undergo a phase
transition to become a strange star. Although we have no direct
experimental evidence which may support the existence of such
objects, recent observational data on the compactness (mass-radius
ratio) of some of the compact objects like Her X-1,
 SAX J1808.4-3658, RX J185635-3754, 4U 1820-30, 4U 1728-34, PSR 0943+10,
strongly favour the possibility that they could actually be strange
stars \citep{Bombaci, Li01, Li02, Li03, Dey01, Xu01, Xu02, Pons}.
Naturally, the internal composition and consequent geometry of such
objects have become a subject of considerable interest over the last
couple of decades. As the physics of very high density matter is
still not very clear, most of the strange star studies have been
performed within the framework of a bag model, see e.g.,
\citet{Glendenning}, \citet{Kettner}. In the phenomenological MIT
bag model \citep{Chodos}, it is assumed that the quark confinement
is caused by a universal pressure $B$ on the surface of any region
containing quarks.

In the bag model, the strange matter equation of state (EOS)
has a simple linear form given by $p = 1/3(\rho - 4B)$ \citep{Witten},
where, $\rho$ is the energy density, $p$ is the isotropic pressure
and $B$ is the bag constant. \citet{Dey01} developed a new model for
strange stars where the quark interaction is described by an interquark
vector potential originating from gluon exchange and a density dependent
scalar potential which restores chiral symmetry at a high density.
The EOS formulated by \citet{Dey01} can also be approximated to a
linear form $p = n(\rho - \rho_{s})$, where, $n$ and $\rho_{s}$ are
two parameters \citep{Gondek, Zdunik}. Thus, if pulsars are actually
strange stars, linearity in EOS seems to be a feature in the composition
of such objects. Once the EOS is known, for a given central density
or pressure, the Tolman-Oppenheimer-Volkoff (TOV) equations are
then integrated to calculate the macroscopic features such as
mass and radius of the star.

However, as the densities within such stars stars are normally
beyond nuclear matter density, one expects anisotropy to play a major
role in such calculations as the TOV equations get modified in
the presence of anisotropic pressures. Note that among many possibilities,
one of the reasons for the consideration of anisotropy within such stars
could be the presence of strong electric field as recently suggested by \citet{Usov}.

Since the initial paper by \citet{Bowers} there has been extensive
research in the study of anisotropic relativistic matter in general
relativity. The analysis of static spherically symmetric anisotropic
fluid spheres is important in relativistic astrophysics. The
investigations of \citet{Ruderman} indicated that nuclear matter may
be anisotropic in high density ranges of order
10$^{15}$~gm~cm$^{-3}$ where nuclear interactions have to be treated
relativistically. Celestrial bodies are not composed of perfect
fluids only: the radial pressure is not equal to the tangential
pressure. \cite{Kippenhahn} showed that anisotropy could be
introduced by the existence of a solid core or the presence of type
3A superfluid. Anisotropy can arise from different kinds of phase
transitions \citep{Sokolov} or pion condensation \citep{Sawyer}.
Anisotropies in spherical galaxies, in the context of Newtonian
gravity, have been studied by \cite{Binney}. The effects of slow
rotation in a star \citep{Herrera} or the mixture of two gases
\citep{Letelier} such as ionized hydrogen and electrons, can be
modelled as a relativistic anisotropic fluid. Strong magnetic fields
can generate an anisotropic pressure component inside a compact
sphere as pointed out by \citet{Weber}.

Our analysis depends on two key assumptions. Firstly, we choose a
rational functional form for the mass function. The form chosen
ensures that the mass function is continuous and well-behaved in the
interior of the star; it yields a monotonically decreasing energy
density in the stellar interior. This is a desirable feature for the
model on physical grounds. It is interesting to observe that many of
the solutions found previously \citep{Chaisi01, Chaisi02} do not
share this feature. It is important to note that the mass function
chosen is similar to the profile that appears in the modelling of
dark energy stars \citep{Lobo}. Secondly, we choose an equation of
state which linearly relates the radial pressure to the energy
density. \cite{Maharaj} have demonstrated that this simple equation
of state is consistent in the modelling of compact matter such as
neutron stars and quasi-stellar objects. The equation of state
yields values of surface redshifts and masses that correspond to
realistic stellar objects such as Her X-1 and Vela X-1
\citep{Gokhroo, Chaisi01}. Linear equations of state are relevant in
the modelling of static spherically symmetric anisotropic quark
matter distributions as demonstrated by \citet{Mak01} and
anisotropic quark stars admitting a conformal symmetry
\citep{Mak03}.

We attempt here to construct  a new model for compact stars where
pressure anisotropy is present and the EOS is linear.
Our paper is organized as follows: In section $2$, we develop a model
for anisotropic stars with an in-built linear EOS. The physical analysis
of the model will be discussed in section $3$. We also perform some numerical
calculations to show the role of anisotropy on the gross features of compact stars.
In section $4$, we summarize our results and point out avenues for further investigation.

\section{Anisotropic model}

We assume the line element for a static spherically symmetric
anisotropic star in the standard form
\begin{equation}
ds^{2} = -e^{\gamma (r)}dt^{2}+e^{\mu (r)}dr^{2}+r^{2}(d\theta ^2 +\sin^2\theta d\phi^2)
\label{mt}
\end{equation}
where $\gamma(r)$ and $\mu(r)$ are the two unknown metric functions.
Assuming the energy momentum tensor for an anisotropic star in the most
general form
\begin{equation}
T_{ij}= \mbox{diag}~(-\rho,~ p_{r},~ p_{t},~p_{t})
\label{em}
\end{equation}
the field equations are obtained as
\begin{eqnarray}
\rho &=& \frac{\left(1-e^{-\mu}\right)}{r^2}+\frac{\mu'e^{-\mu}}{r} \label{en}\\
p_{r} &=& \frac{\gamma'e^{-\mu}}{r}-\frac{\left(1-e^{-\mu}\right)}{r^2}\label{rp}\\
p_{t} &=& \frac{e^{-\mu}}{4}\left(2\gamma''+{\gamma'}^2-\gamma'\mu'+\frac{2\gamma'}{r}-
\frac{2\mu'}{r}\right)\label{tp}
\end{eqnarray}
where primes denote differentiation with respect to $r$. In equations~(\ref{en})-(\ref{tp}), $\rho$ is the energy density, $p_{r}$ is the radial pressure and $p_{t}$ is the tangential pressure. These equations may be combined to yield
\begin{equation}
(\rho+p_{r})\gamma' + 2p_{r}'+\frac{4}{r}(p_{r} - p_{t}) = 0\label{con}
\end{equation}
which is the conservation equation. If the mass contained within a radius $r$ of the star is defined as
\begin{equation}
m(r) = \frac{1}{2} \int_0^r r^{\prime 2} \rho (r^{\prime})\label{ms}
dr^{\prime}
\end{equation}
we obtain the equivalent system of field equations
\begin{eqnarray}
e^{-\mu} &=& 1-\frac{2m(r)}{r}\label{m1}\\
\gamma' &=& \frac{2m(r) + p_{r} r^3}{r(r-2m)}\label{m2}\\
\frac{dp_{r}}{dr} &=& - \left(\rho + p_{r}\right)\frac{2 m(r)
+ r^3 p_{r}}{2r(r-2m(r))} + \frac{2\Delta}{r}\label{mtov}
\end{eqnarray}
where $\Delta = p_{t} - p_{r}$, is the measure of anisotropy in
this model. Equation~(\ref{mtov}) is the modified TOV equation
in the presence of anisotropy.

To make the above set of equations tractable, we choose the mass
function in a particular form
\begin{equation}
m(r) = \frac{b r^3}{2(1+a r^2)}\label{mas}
\end{equation}
where $a$ and $b$ are two positive constants. The choice of the mass
function is motivated by the fact that it gives a monotonically
decreasing energy density in the stellar interior. Similar forms for
the mass function have earlier been considered by \citet{Matese} and
\citet{Finch} for isotropic fluid spheres, \citet{Lobo} for dark
energy stars, and \citet{Mak02} for anisotropic fluid spheres.

Substituting equation~(\ref{mas}) in equation~(\ref{m1}), we obtain
one of the metric functions in the form
\begin{equation}
e^{\mu} = \frac{1+ar^2}{1+(a-b)r^2}\label{mt1}.
\end{equation}
From equation~(\ref{en}), the energy density then can be obtained as
\begin{equation}
\rho = \frac{b(3+ar^2)}{(1+ar^2)^2}\label{den}
\end{equation}
so that the central density is given by $\rho_{c} = 3b$.

To find the other metric function $\gamma(r)$, in general,
either a special functional form of the anisotropic
parameter $\Delta (r)$ (see \citet{Mak02}, \citet{Dev}) or
the pressure $p_{r}(r)$ profile (see \citet{Chaisi01}) is
chosen at this point. However, since linearity seems
to be a feature in compact star EOS (as shown by \citet{Sharma01, Sharma02}), we
assume a linear form of the EOS to solve equation~(\ref{m2}).
Integrating equation~(\ref{m2}) we get
\begin{equation}
\gamma = \int{\frac{p_{r} (1+ar^2)r}{1+(a-b) r^2}}dr
+\frac{b}{2(a-b)} \ln(1+ar^2-br^2) +\ln B \label{m3}
\end{equation}
where, $\ln B$ is an integration constant. To evaluate the
first term on right hand side of equation~(\ref{m3}),
we assume a linear EOS of the form
\begin{equation}
p_{r} = n(\rho - \rho_{s}),
\end{equation}
where $0 \leq n \leq 1$ is a constant which is related to
the sound speed $dp_{r}/d\rho=n$, and $\rho_{s}$ is the
density at the surface $r=s$, where the radial pressure vanishes.
We do not impose any restrictions on the tangential pressure $p_{t}$.
We obtain the other metric function in the form
\begin{equation}
e^{\gamma} = B (1+ar^2)^n (1+ar^2-br^2)^k \exp\left[-\frac{acr^2}{2(a
-b)}\right]\label{soln}
\end{equation}
where
\begin{equation}
k=\frac{5nab-2na^2-3nb^2+ab-b^2-bc}{2(a-b)^2}\label{fac}.
\end{equation}
In equation~(\ref{fac}) we have set $c=n\rho_{s}$.
Finally using equation~(\ref{mtov}), we get the anisotropic parameter as
\begin{equation}
\Delta = \frac{r^2(\rho+p_{r})}{4(1+ar^2-br^2)}[p_{r}(1+ar^2)+b]
- \frac{abnr^2(5+ar^2)}{(1+ar^2)^3}\label{anp}
\end{equation}
which clearly shows that anisotropy vanishes at the
centre in this model, i.e., $p_{r}=p_{t}$ at $r=0$. The metric potentials
are nonsingular at the centre. The energy density and the two pressures ($p_{r}$ and $p_{t}$) are also well behaved in the stellar interior as will be shown later.

In our model, $a$, $b$, $c$ have the dimension of $length^{-2}$.
For simplicity in numerical calculations, we make the following transformations:
$$\tilde{a} = a R^2,~~~ \tilde{b} = b R^2,~~~ \tilde{c} = c R^2,$$
where, $R$ is a parameter which has the dimension of a $length$. In
terms of these dimensionless parameters our results may be summarized as follows:
\begin{eqnarray}
e^{\mu} &=& \frac{1+\tilde{a}y}{1+(\tilde{a} -\tilde{b})y}\\
e^{\gamma} &=& B(1+\tilde{a}y)^n[1+(\tilde{a}-\tilde{b})y]^{\tilde{k}}
\exp\left[-\frac{\tilde{a}\tilde{c}y}{2(\tilde{a}-\tilde{b})}\right]\\
\rho &=& \frac{\tilde{b}(3+\tilde{a} y)}{R^2(1+\tilde{a}y)^2}\\
p_{r} &=& n(\rho - \rho_{s})\\
p_{t} &=& p_{r} +\Delta
\end{eqnarray}
where
\begin{equation}
\tilde{k}  = \frac{5n\tilde{a}\tilde{b}-2n\tilde{a}^2-
3n\tilde{b}^2+\tilde{a}\tilde{b}-\tilde{b}^2-\tilde{b}\tilde{c}}{2(\tilde{a}-\tilde{b})^2}
\end{equation}
and $y=r^2/R^2$.
We may also write the anisotropic parameter in the form
\begin{eqnarray}
\Delta &=& \frac{r^2}{4R^4(1+\tilde{a} y -\tilde{b} y)(1+\tilde{a}y)^3} \times  \nonumber  \\
& & [n(n+1)\tilde{b}^2(3+\tilde{a}y)^2- (2n+1)\tilde{b}\tilde{c}(3+\tilde{a}y)(1+\tilde{a}y)^2  \nonumber \\
& & -\tilde{c}^2(1+\tilde{a}y)^4
+\tilde{b}^2(n+1)(3+\tilde{a}y)(1+\tilde{a}y) \nonumber \\
& & -\tilde{b}\tilde{c}(1+\tilde{a}y)^3 - 4n\tilde{a}\tilde{b}(5+\tilde{a}y)(1+\tilde{a}y -\tilde{b}y)]\label{delt}.
\end{eqnarray}
The mass contained within a radius $s$ is given by
\begin{equation}
M = \frac{\tilde{b}s^3/R^2}{2(1+\tilde{a}s^2/R^2)}\label{tm}.
\end{equation}

\section{Physical analysis}

In this section, following \citet{Delgaty}, we impose some restrictions on the model to make it physically acceptable.
\begin{itemize}
\item The interior solution should be matched with the Schwarzschild exterior model at the boundary $r=s$, which gives
\begin{eqnarray}
e^{\mu(r=s)} &=& \left(1-\frac{2M}{s}\right)^{-1}\label{se1} \\
e^{\gamma(r=s)} &=& \left(1-\frac{2M}{s}\right)\label{se2}.
\end{eqnarray}
\item  The values of $\tilde{a}$, $\tilde{b}$ should be so chosen that the energy density $\rho$ and the radial pressure $p_{r}$ remain positive inside the star.
\item The causality condition is satisfied if we restrict our model with $0 \leq n \leq 1$.
\item The value of $\Delta$ should be so chosen that $\frac{dp_{r}}{dr} \leq 0$ in the interior of the star and the tangential pressure $p_{t} \geq 0$. This imposes a bound on the anisotropy given by
\begin{equation}
\Delta \leq  \frac{r^2(\rho+p_{r})}{4(1+(\tilde{a}-\tilde{b})r^2/R^2)}[p_{r}(1+\tilde{a}r^2/R^2)+\tilde{b}/R^2]\label{alt}
\end{equation}

\item The model has five independent parameters $\tilde{a}$, $\tilde{b}$, $\tilde{c}$, $n$ and $B$. Note that $R$ is not independent as it can be expressed in terms of $\tilde{a}$, $\tilde{b}$ and $\tilde{c}$. Two of the above parameters can be fixed by the matching conditions given by equations~(\ref{se1}) and (\ref{se2}). The third parameter becomes fixed if we choose the central/surface density. Thus we will be left with two free parameters governing the geometry and EOS of the star.
\end{itemize}

By suitably choosing values of the unknown parameters, it is
possible to show that our model can describe realistic compact
stellar objects. For example, let us consider a particular case
where the central density and the surface density are given by
$\rho_{c}=4.68\times10^{15}$~gm~cm$^{-3}$ and
$\rho_{s}=1.17\times10^{15}$~gm~cm$^{-3}$, respectively. If we now
choose, $\tilde{a}=53.34$, $\tilde{b}=54.34$ and $R=43.245$~km, then
the model yields a star of mass $M=1.435~M_{\odot}$ and radius
$s=7.07$~km. Note that these are the values which can be obtained if
we use one of the EOS for strange matter formulated by
\citet{Dey01}. The behaviour of the energy density, two pressures
and anisotropic parameter are shown in Figures 1-3, respectively. It
is to be noted here that, unlike the slope in \citet{Dey01} (which
was $n=0.463$), the maximum slope for a physically reasonable model
in this case is given by $n=0.297$. This suggests that if we want to
obtain the same mass and radius for the central and surface
densities as in \citet{Dey01} (which arguably described the X-ray
pulsar SAX J 1808.4-3658), and include anisotropy which vanishes at
the boundary, then the EOS becomes much softer, i.e. the gradient
$dp_{r}/d\rho$ decreases. On the other hand, keeping the same slope
($n=0.463$) and surface density ($\rho_{s} = 1.17\times
10^{15}$~gm~cm$^{-3}$), will give us a star of mass $\sim
1.96~M_{\odot}$. Choosing different sets of values will obviously
give different results as shown in Table 1. Thus anisotropy provides
a tool to consider a large class of degenerate states in this model.

\begin{table}
\centering
\caption{Central density and mass for different anisotropic stellar models. We have set, $s=7.07$~km, $ \rho_s = 1.17\times 10^{15}$~gm~cm$^{-3}$ and $R=43.245$~km.}
\begin{tabular}{@{}|l|c|c|c|r|@{}}   \hline
  $\tilde{b}$ & $\tilde{a}$ & $n$ & $\rho_{c}$ &  Mass \\
   & & & (gm~cm$^{-3}$) & ($M_{\odot}$)    \\ \hline
 30 & 23.681 & 0.271 & 2.58 & 1.177 \\
 40 & 36.346 & 0.277 & 3.44 &  1.300 \\
 50 & 48.308 & 0.290 & 4.31 &  1.398 \\
 53.34 & 54.340 & 0.297 & 4.68 & 1.435 \\
 60 & 59.789 & 0.306 & 5.17 &  1.479 \\
 70 & 70.921 & 0.322 & 6.03 &  1.548 \\
 80 & 81.786 & 0.337 & 6.89 &  1.608 \\
 90 & 92.442 & 0.353 & 7.75 &  1.661 \\
 100 & 102.929 & 0.367 & 8.61 &  1.707 \\
 183 & 186.164 & 0.463 & 15.76 & 1.962 \\ \hline
\end{tabular}
\end{table}

\section{Discussion}

We have obtained new exact solutions to the Einstein field equations
capable of describing relativistic compact objects with anisotropic
matter distribution and admitting a linear EOS. The solutions
obtained are nonsingular and well behaved in the stellar interior.
Unlike other models \citep{Mak02, Dev}, the form of the anisotropic
parameter is not chosen a priori in this model. We have shown that
the consideration of anisotropy may provide a wide range of
consequences in the geometry and composition of a star. For example,
this may either yield a softer EOS (so that the gradient
$dp_{r}/d\rho$ may be different) or a star with different mass and
radius. Our analysis depends critically on the choice of the mass
function given by equation~(\ref{mas}) so that a linear equation of
state is possible. Note that some other treatments, such as the
result of \citet{Maharaj}, with a linear EOS has the unrealistic feature
of vanishing energy density at the boundary. Our model does not
suffer from this defect. It is also important to observe that in the
presence of anisotropy the material composition, in objects such as
SAX J1808.4-3658, need not necessarily be quark matter because the softness may vary.
In future work it would be interesting to investigate what other
forms of the mass function are consistent with a linear equation of
state.

\section*{Acknowledgment}
RS acknowledges the financial support (grant no. SFP2005070600007)
from the National Research Foundation (NRF), South Africa. SDM acknowledges
that this work is based upon research supported by the South African Research
Chair Initiative of the Department of Science and Technology and the National
Research Foundation.

\begin{figure}
\centering
\includegraphics[scale=0.8]{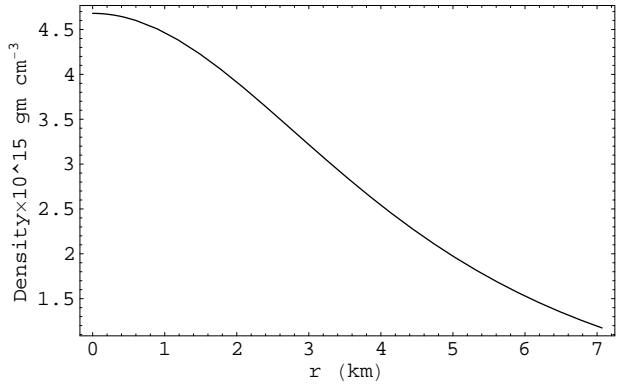}
\caption{\label{fg1}Energy density plotted against radial distance.}
\end{figure}
\begin{figure}
\centering
\includegraphics[scale=0.8]{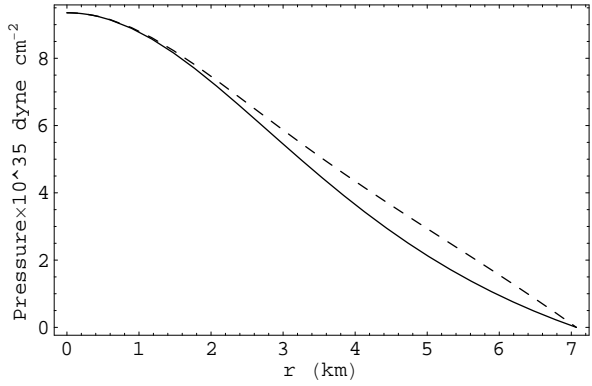}
\caption{\label{fg2} Radial pressure (solid line) and tangential
pressure (dashed line) plotted against distance.}
\end{figure}
\begin{figure}
\centering
\includegraphics[scale=0.8]{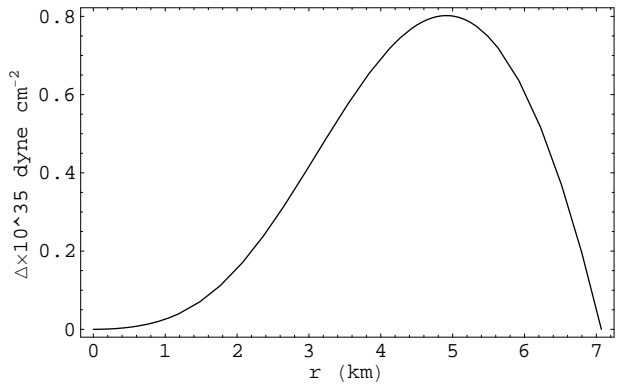}
\caption{\label{fg3}Anisotropic parameter plotted against radial distance.}
\end{figure}

\label{lastpage}

\end{document}